\documentclass[proof]{pasj00}     
\draft
\SetRunningHead{Y. Abe et al.}{Three Spectral State of 4U 1630-47}
\Received{2004 November 17}
\Accepted{2005 May 10}
\Published{出版日}         
\begin{document}
\title{Three Spectral States of the Disk X-Ray Emission of the Black-Hole
Candidate 4U 1630-47} 

\author{Yukiko \textsc{Abe},\altaffilmark{1}  
        Yasushi \textsc{Fukazawa},\altaffilmark{1} 
        Aya \textsc{Kubota},\altaffilmark{2}
        Daisuke \textsc{Kasama},\altaffilmark{3} \\
        and 
        Kazuo \textsc{Makishima}\altaffilmark{2,3}}
\altaffiltext{1}{Department of Physical Science, School of Science,
Hiroshima University, 1-3-1 Kagamiyama, \\ 
 Higashi-Hiroshima, Hiroshima 739-8526}
\email{abe@hirax7.hepl.hiroshima-u.ac.jp, fukazawa@hirax6.hepl.hiroshima-u.ac.jp}
\altaffiltext{2}{The Institute of Physical and Chemical Research (RIKEN), Wako, Saitama, 350-0198}
\altaffiltext{3}{Department of Physics, School of Science, The University 
of Tokyo, 7-3-1 Hongo, Bunkyo-ku, Tokyo 113-0033}

\KeyWords{ accretion, accretion disks ---  black hole physics ---
stars: individual (4U 1630-47) ---  X-rays: stars}

\maketitle

\begin{abstract}
We studied a time history of X-ray spectral states of a black-hole candidate,
4U 1630-47, utilizing data from a number of monitoring observations
with the Rossi 
X-Ray Timing Explorer over 1996--2004. 
These observations covered five outbursts of 4U 1630-47, 
and most of the data recorded typical features of the high/soft states. 
We found that the spectra in the high/soft states can be further classified
into three states. 
The first spectral state is explained by a concept of the 
standard accretion disk picture.
The second state appears in the so-called very high state, 
where a dominant hard component is seen and the disk radius 
apparently becomes too small.
These phenomena are explained 
by the effect of inverse Compton scattering of disk photons, 
as shown by Kubota, Makishima, \& Ebisawa (2001, ApJ, 560, L147) 
for GRO J1655-40.
The third state is characterized in such a way that the disk luminosity
 varies in proportion to $T_{\rm in}^2$, rather than $T_{\rm in}^4$,
 where $T_{\rm in}$ is the inner-disk temperature. 
This state is suggested to be an optically-thick and advection-dominated slim
disk, as suggested by Kubota \& Makishima (2004, ApJ, 601, 428) 
for XTE J1550-564.
The second and third states appear, with good reproducibility, when $T_{\rm in}$ and the total X-ray luminosity are 
higher than 1.2 keV and $\sim2.5\times10^{38}\left(D/10\quad{\rm kpc}\right)^2\left[\cos{\theta}/\left(1/\sqrt{3}\right)\right]^{-1}$ 
erg s$^{-1}$, respectively, where $D$ is the distance to the object and
 $\theta$ is the inclination angle to the disk. 
The present results suggest that these three spectral states 
commonly appear among black-hole binaries under high accretion rates.
\end{abstract}

\section{Introduction}
The physics of accreting black holes is naturally the same among 
stellar-mass black holes, ultraluminous X-ray sources (ULXs), and active
galactic nuclei (AGN), 
and information concerning accretion disks enables us to understand the 
spacetime geometry around the central black holes.
Among many classes of black holes, 
galactic black hole binaries (hereafter BHBs) are the best for
detailed spectral and timing studies, 
and often the observational results of BHBs are considered as a 
basis to understand other classes of black holes. 
Therefore, it is essentially important to precisely understand the 
accretion-disk structure of BHBs from observed spectra.

In classical understanding, the X-ray spectrum of BHBs 
exhibits two distinct features of high/soft and low/hard states 
(e.g. \cite{Tanaka}). 
In the high/soft state, the spectrum consists of a dominant soft component accompanied by a powerlaw
component with a photon index of $\Gamma=2$--2.5 (e.g., \cite{Tanaka}). 
The dominant soft component is believed to be a thermal emission from 
a geometrically thin and optically thick standard accretion disk 
(\cite{Shakura}) extending down to the 
last Keplerian orbit around the central black hole.
This is the standard view of the high/soft state BHBs. 
In fact, this soft component is well reproduced by a multicolor disk
model (MCD model or {\it diskbb} in the {\it xspec} package; \cite{Mitsuda}) 
that approximates the integrated emission from the standard disk. 
The MCD model has two parameters: an inner disk temperature $T_{\rm in}$ 
and a normalization 
$K_{\rm MCD} =r_{\rm in}^2/(D/10\quad{\rm kpc})^2\cos \theta$
with an apparent disk inner radius $r_{\rm in}$, a
disk inclination angle $\theta$, and a source distance $D$.
Correcting $r_{\rm in}$ with the 
inner boundary condition (\cite{Kubota1998}) and color-to-effective 
temperature, we can estimate the true inner radius, $R_{\rm in}$. 
Past observations of many BHBs revealed that $R_{\rm in}$ 
is kept remarkably constant even though the X-ray luminosity varies 
considerably (\cite{Makishima1986, Takizawa, Ebisawa1993}, 1994). 
Moreover, through studies of many BHBs, including Cyg~X-1
(\cite{Dotani}), 
the inner radius $R_{\rm in}$ is found to be consistent with 
the last stable orbit, $R_{\rm so}=3R_{\rm g}=6GM/c^2$, where 
$R_{\rm g}$ is the Schwarzschild radius, $G$ is the gravitational
constant, $c$ is the light velocity, and
$M$ is the mass of the non-spinning black hole.
Therefore, it is strongly supported that the standard view is realized in the high/soft state BHBs. 

However, this simple, but physically well-understood, picture was found to be 
sometimes broken when the disk luminosity $L_{\rm disk}$ 
reaches a certain critical luminosity.
The biggest problem is that the estimated values of $R_{\rm in}$ do not
remain constant when the powerlaw contribution becomes significant
for the total luminosity.
Based on the Rossi X-Ray Timing Explorer ({\it RXTE}) 
observations of the galactic BHBs GRO~J$1655-40$ 
and XTE~J$1550-564$,
\citet{Kubota2001}, \citet{Kubota2004} found that this anomaly 
is observed when the disk luminosity exceeds a certain critical value
(or a critical temperature around 1.0--1.2 keV), while
the standard view is realized below such a luminosity.
They called this branch an {\it anomalous regime} to distinguish from the 
standard high/soft state ({\it standard regime}). 
Together with the anomalies in $R_{\rm in}$,  the {\it anomalous regime} 
can be characterized by a 
large intensity variation associated with quasi-periodic oscillations
(QPOs) and the strong hard emission described by steeper powerlaw 
of $\Gamma=2.5$--3. 
The characteristics of the {\it anomalous regime} are naturally the same as 
those of the very high state in the literatures 
(e.g., \cite{Miyamoto}; \cite{McClintock}; \cite{van der Klis}).
\citet{Kubota2001}, \citet{Kubota2004} found 
that the strong hard emission is caused 
by inverse Compton scattering due to high-energy electrons that may 
reside around the disk, and for the first time found quantitatively 
that the $R_{\rm in}$ returns to
the same value as seen in the {\it standard regime}, if the Compton effect
is taken into account. 
This scenario is consistent with those of QPOs, which are usually 
thought to be related to the corona surrounding the disk.

\citet{Kubota2001}, \citet{Kubota2004} also found that another unusual state 
appeared in the brightest period of XTE J1550-564 and the beginning of the 
outburst of GRO~J$1655-40$.
The X-ray spectrum is apparently expressed by the dominant soft
component and very weak powerlaw
tail like in the case of the standard high/soft state. 
However,
$L_{\rm disk}$ is not proportional to $T_{\rm in}^4$, but to 
$T_{\rm in}^2$ (or $r_{\rm in}\propto T_{\rm in}^{-1}$) under the MCD 
plus powerlaw model,
and the spectral shape of the soft component is somewhat distorted from
that of the standard disk.
This regime is called an {\it apparently standard regime} by \citet{Kubota2004}.
They suggested that the {\it apparently standard
regime} shows a sign of a slim disk, in which optically thick advective
cooling becomes significant (\cite{Abramowicz}).

These observed three spectral regimes seem to be consistent with 
the theoretically predicted $S$-shape sequence for the optically thick 
accretion-disk solutions.
This correspondence is simple, and 
it is sometimes referred to understand the spectral behavior of ULXs 
(e.g., \cite{Kubota2002}; Mizuno et al. 2001).
However, such three regimes have ever been confirmed for only two Galactic
black holes, GRO J1655-40 and XTE J1550-564, 
and it is not yet clear whether these regimes are general.
Moreover, we have not known whether the critical luminosity is always the same 
in recurrent BHBs that repeat outbursts.
In this paper, we hence present spectral analyses of a
recurrent black hole binary, 4U~1630-47.

This object is one of the famous historical BHBs (e.g., \cite{Tanaka}), 
and is known to repeat X-ray outbursts
with a period of about 600--650 days (\cite{Jones}; Parmar et al. 1995).
No optical counterpart is known for 4U~1630-47, 
probably due to its high reddening and crowded field 
(Parmar et al. 1986). 
A number of authors have shown the observational results of this source 
based on X-ray observations with
{\it Ginga} (\cite{Parmar1997}), 
{\it EXOSAT} (Parmar et al. 1986; Kuulkers et al. 1997), 
{\it BATSE/CGRO} (\cite{Bloser}), 
{\it Beppo-SAX} (\cite{Oosterbroek}), and so on.
We here analyzed 322 {\it RXTE} data of this source during 1996--2004 
observations, consisting of 5 outbursts.
{\it RXTE} data of the 1996 and 1998 outbursts 
have already been analyzed by other authors, including 
\citet{Tomsick1998}, \citet{Kuulkers1998}, \citet{Cui}, \citet{Hjellming}, \citet{Dieters}, \citet{Trudolyubov}, and \citet{Kalemci}.
The anomaly of $R_{\rm in}$ (e.g., \cite{Oosterbroek}), together with 
existence of QPOs (e.g., \cite{Tomsick}) during the 1998 outburst
has been reported.

The aim of the present work was (1) to test whether 
the {\it anomalous regime} and {\it apparently standard regime} are 
found in the complete {\it RXTE} data set of this source, 
(2) to determine the critical luminosity and to test whether the
critical luminosity is the same among all outbursts of this single source, 
(3) to confirm the scenario of the inverse Compton scattering 
in the {\it anomalous regime}
(i.e.,  to test whether the scenario 
can solve the problem of variable $R_{\rm in}$), 
and (4) to test the slim disk scenario if the {\it apparently standard 
regime} appears, and 
then we can discuss the generality of the scenario suggested by 
\citet{Kubota2001} and \citet{Kubota2004}. 
This is very important to understand the spectral evolution of 
accreting black holes.
The black hole mass, distance $D$, and inclination $\theta$ of 4U
1630-47 are not
determined; in this paper we assume
$D=10$ kpc and $\cos{\theta}$ = $1/\sqrt{3}$, following the previous
works, and do not treat the data in the low state.

Before presenting the analyses, we briefly compare the spectral regimes
applied in this paper to 
a more seminal definition of the spectral and timing behavior used in other 
literature in section 2. 
In section 3, 
we show the observation and data-reduction criteria, and in subsection 4.1, 
we characterize the observed spectral states based on the canonical MCD 
plus powerlaw model.
In subsection 4.2 and subsection 4.3, 
we consider the effects of the inverse Compton effect
and the slim disk, respectively.
In section 5, we give a discussion on the results from the view point of the
disk structure.

\section{Brief Notes On Seminal Definition of Spectral/Timing States}

The spectral/timing states of BHBs are sometimes complicated. 
Therefore, we here briefly show the correspondence of the 
spectral states that we used in the present work to the states defined 
in other literature. 
A summary is given in table~1.
Beyond the two classical low and high states, 
several subdivisions are defined in the literature:
the very high state and/or the intermediate state, 
the disk dominant high/soft state, the low/hard, 
and the off/quiescent state (e.g., \cite{McClintock}). 

The off/quiescent state is in an extraordinarily faint period
($L_{\rm X}\simeq 10^{30.5-33.5}~{\rm erg~s^{-1}}$) and 
generally understood by a picture of advection-dominated accretion flow 
(ADAF: \cite{Narayan1994}, 1995; Narayan et al. 1996).
The low/hard state is characterized by a single powerlaw 
spectrum of $\sim 1.7$ with an exponential cutoff 
around several tens of keV; 
the accretion is thought to still be dominated by advection.
Large intensity variation of a strong flat-top noise
 accompanied by type-C (or C') QPOs with a variable centroid 
frequency (\cite{Remillard}) 
is usually observed in this state (\cite{McClintock}). 

The disk-dominant high/soft state is just the same as the {\it standard 
regime}, where the standard disk is realized. 
The intensity variation is very weak and QPOs are seldom observed, 
except on rare occasion of type-A QPOs associated with a weak red noise 
(e.g., \cite{Remillard, Nespoli, Casella, Homan}). 
Based on the spectral shape and the luminosity, the {\it apparently standard
regime} is also classified into the disk-dominant high/soft state, 
even though we are suggesting that the standard disk is not
justified in this regime.

\begin{table}
\centering
\begin{minipage}{175mm}
\caption{Correspondence of the definition of the states in this paper 
to those in other papers.}
\label{tab:name}
\begin{tabular}{p{2.5cm}cccc}
\hline \hline
this paper & classification & MR03$^{\ast}$ & QPO type $^\dagger$ & Remarks $^\ddagger$\\
\hline
(off/quiescent)  &off/quiescent & hard & unknown & thermal IC \\
(low/hard)  &low/hard & hard &C or C' &   thermal IC\\
(very high state with strong hard emission) & very high (intermediate) & SPL & C or C' &thermal IC\\
anomalous  &  very high (intermediate) & SPL & B & thermal IC\\
standard  &high/soft  & TD & A or none &disk dominant + PL\\
apparently standard  & high/soft & TD & none & disk dominant + weak PL\\
\hline
\multicolumn{5}{p{16cm}}{$^\ast$ State definition by McClintock and Remillard
 (2003). SPL and TD represents a steep powerlaw state and a thermal
 dominant state, respectively.}\\
\multicolumn{5}{p{16cm}}{$^\dagger$ QPO type is referred to Remillard et
 al. (2002). Note that majority of QPOs are found as type B.}\\
\multicolumn{5}{p{16cm}}{$^\ddagger$ Characteristics of X-ray spectrum. IC and
 PL represent an inverse Compton scattering model and a powerlaw model, 
respectively}\\
\end{tabular}
\end{minipage}
\end{table}

Definitions of the very high state and/or the intermediate state are 
somehow controversial.
In the original paper by \citet{Miyamoto}, 
the very high state was first defined as a high luminosity state 
with a large intensity variation, 
in contradiction to the standard high/soft state.
After their pioneering work, 
spectral studies revealed that the very high state 
is associated with a very strong hard component with a steeper photon 
index of $\Gamma=2.4$--3.0 than the hard tail in the standard high/soft 
state (\cite{McClintock} and references therein).
Based on this strong and steep hard component, 
\citet{McClintock} called the very high state the steep powerlaw state.
Detailed timing studies found that, in the very high state, QPOs 
were generally observed like in the case of the low/hard 
state (\cite{van der Klis95}, 2004 and references therein).
In the case of XTE~J1550-564, 
type-B QPOs are observed when the hard emission is less dominant 
(but still much stronger than the standard high/soft state), 
while type-C (or C') QPOs are observed when the hard component is the 
most significant in the total emission 
(e.g., \cite{Remillard, Kubota2004}). 
Some authors suggest that the very high state is itself the 
intermediate state of transition between the low/hard state 
and the disk-dominant high/soft state 
(e.g., \cite{Rutledge, Homan, van der Klis95}), 
while other authors 
suggest that the very high state is a bona fide state of BHBs, and 
note that nothing in their definitions constrains the order in which 
state transitions should occur (e.g., \cite{McClintock}).
Recently, \citet{belloni05} classified the very high state and/or
intermediate state into two regimes along the flare evolution.
As previously described, the {\it anomalous regime} is basically the
same as the very high state, but strictly speaking, corresponds to the 
very high state with type-B QPOs and weaker hard emission
(\cite{Kubota2004d}) or the soft intermediate state (\cite{belloni05}).
The very high state with much stronger hard emission 
accompanied by type-C QPOs was found just after the low/hard state, 
would correspond to the hard intermediate state (\cite{belloni05}), 
and the optically thick disk is somehow truncated 
at $>R_{\rm so}$ (\cite{Kubota2004d}). 
Thus, this type of the very high state may be associated with the state 
transition.

Thus, the states of bright BHBs are still being discussed from the view
point of both spectral and timing phenomena.
On the other hand, three regimes treated in the present work are based on
the spectral features of the disk emission, 
including the standard disk model, the inverse Compton scattering, 
and the slim disk model.
In this paper, to understand the disk structure in the high state,
we focus on whether the standard disk picture is valid or broken 
against the luminosity and disk temperature, by considering the effects
of the inverse Compton scattering and the slim disk, rather than
reclassifying the high states.
Such an approach is important to describe many complex features found in
the bright period of BHBs, not phenomelogically, but physically.

\section{Observation and Data Reduction}
\begin{figure}[hbt]
\begin{center}
\rotatebox{0}{\resizebox{10cm}{!}{\includegraphics{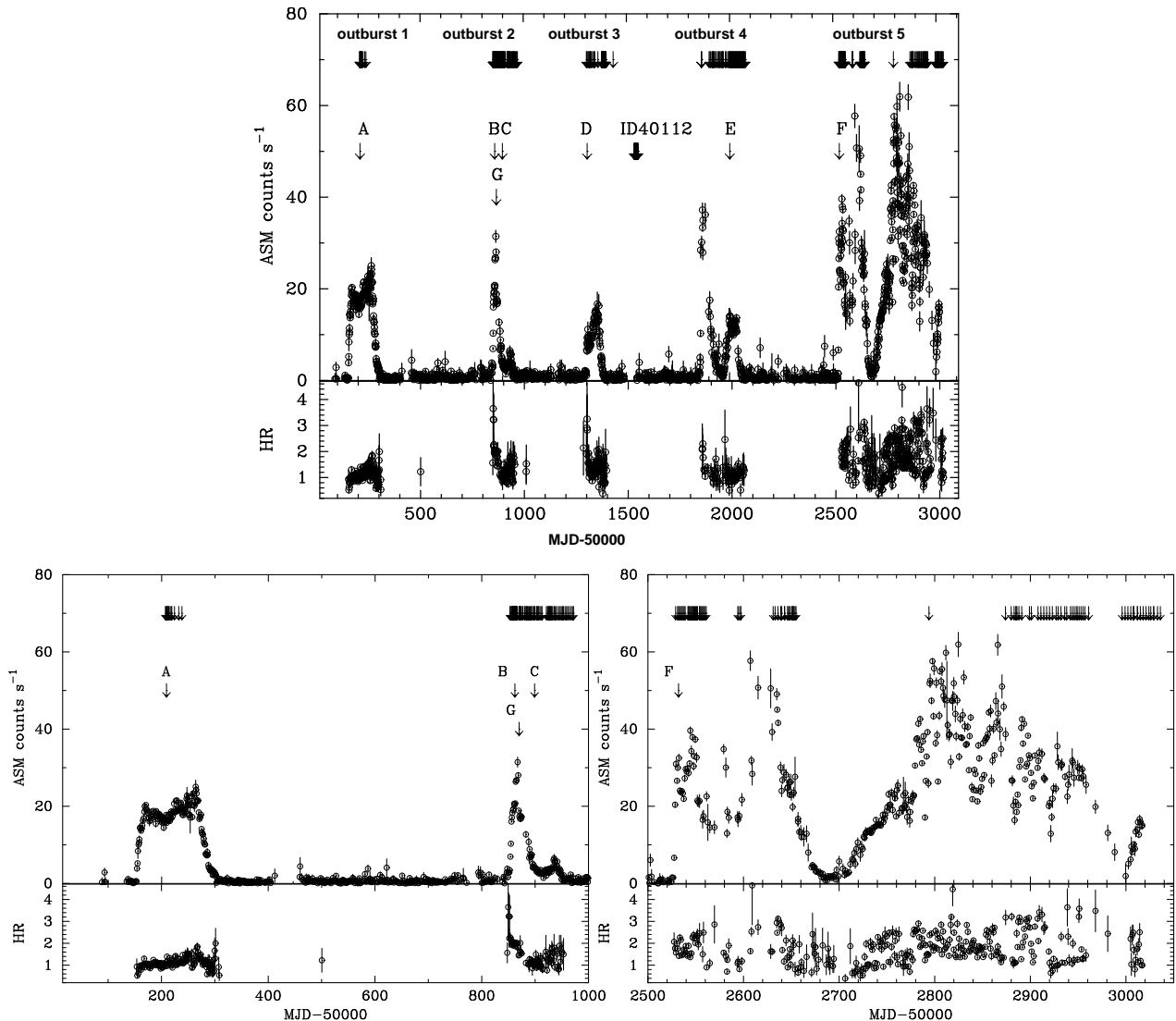}}}
\end{center}
\begin{minipage}[tbhn]{8cm}
\rotatebox{-90}{\resizebox{6.5cm}{!}{\includegraphics{psfile/figure1b.ps}}}
\end{minipage}\quad
\begin{minipage}[tbhn]{8cm}
\rotatebox{-90}{\resizebox{6.5cm}{!}{\includegraphics{psfile/figure1c.ps}}}
\end{minipage}
\caption{The top panel shows time histories of the X-ray count rate and the 
hardness ratio (HR) (5--12 keV/3--5 keV) of 4U 1630-47 by the {\it RXTE}/ASM
 observation in 1996--2004. The PCA pointing observations are indicated
 with arrows at the top. ``A'', ..., ``G'', and ``ID 40112'' indicate the
 observation data IDs referred to in the text.
The bottom panels are enlargements of outbursts 1, 2 (left), and
 5 (right).}
\label{asmlc}
\end{figure}

The black-hole candidate 4U 1630-47 has been continuously monitored by the
{\it RXTE}/ASM (Bradt et al. 1993; \cite{Levine}). 
Figure \ref{asmlc} shows time histories 
of a 1.5--12 keV count rate and a hardness
ratio of this source in 1996--2004, obtained with the {\it RXTE}/ASM. 
The source showed five outbursts since 1996. 
We analyzed 322 pointing observations with the {\it RXTE}/PCA of the five
outbursts, which are indicated with arrows at the top in figure \ref{asmlc}. 
In addition to the PCA data, the HEXTE data were analyzed only when their
data quality was good. 
We analyzed the PCA standard-2 data and the HEXTE standard mode data
for spectral fittings,
and performed data selection on the condition of a minimum elevation
angle of 10$^\circ$ above the Earth's limb. 
Since the spacecraft attitude is not stable just after changing the 
pointing direction, we
excluded the data obtained when the actual pointing direction is more
than 0$\acute{.}$02 away from the aimed direction. 
The selected data of the individual proportional counter units were 
co-added for spectral analyses. 
Systematic errors of 0.3$\%$ were added to the PCA spectra; 
furthermore,
we added 2\% and 10\% systematic errors to the 4--8 keV
and 25--30 keV band, respectively, referring to the PCA calibration results
in \citet{Jahoda}.
The latter is due to the uncertainty associated with 
the instrumental Xenon edge.
The response matrices are generated by the ftool {\it pcarsp},
considering the change of calibration parameters, including the gain of PCA.

We also utilized the ASCA/GIS data (\cite{Ohashi}; \cite{Makishima1996}) 
for constraining the spectra at 
lower energies.
The ASCA observation of this source was performed on 1998 February 26
(MJD = 50870), and this was almost simultaneous with the {\it RXTE}/PCA
observation (observation ID 30188-02-20-00).
The GIS events were
extracted from a circular region of 6$'$ radius centered on the
image peak, after selecting good time intervals in a standard
procedure. 
The dead-time fractions were determined to be 9$\%$ for GIS 2
and 12$\%$ for GIS 3 in reference to the count-rate monitor data 
(\cite{Makishima1996}).

In pointing observations of ID 40112 in figure \ref{asmlc}, 
the emission from 4U 1630-47 was very
 weak, and the count rate was constant, but higher, than the
 instrumental background level of the PCA. 
Considering further that a strong iron-line is present in the spectra, it is thought that diffuse X-ray emission
 from the Galactic plane was mainly observed in this period. 
This emission must be a background for the observed PCA spectra of
4U 1630-47. 
We fitted 29 pointing observations of ID 40112 with a thermal 
bremsstrahlung and a Gaussian representing the iron-line, 
multiplied by the interstellar absorption. Table \ref{par_bgd} gives the
values of the model parameters, each averaged over the 29 spectra. In
a subsequent model fitting to the PCA spectra of 4U 1630-47, we included
the Galactic X-ray background as a fixed model component, using the
parameters in table \ref{par_bgd}.

\begin{table} [!hbt]
\vspace{0.5cm}
\caption{Parameters for the Galactic diffuse background model, obtained
 from the data ID 40112.}
\label{par_bgd}
\begin{center}                           
\begin{tabular}{lccccccc} \hline\hline
model & wabs$^{\ast}$ & & \multicolumn{2}{c}{bremss$^{\ast}$} & & \multicolumn{2}{c}{Gaussian$^{\ast}$}  \\ \cline{2-2}\cline{4-5}\cline{7-8}
      & $N_{\rm H}$ & & $kT$ & Normalization$^{\dagger}$ & & Line $E$ & Normalization \\ 
      & (10$^{22}$cm$^{-2}$) & & (keV) & & & (keV) & (photons
 cm$^{-2}$ s$^{-1}$) \\ \hline
average   & 1.62 & & 17.0 & 0.025 & & 6.58 & $5.98\times10^{-4}$ \\ \hline 
\multicolumn{8}{p{16cm}}{$\ast$: ''wabs'', ``bremss'', and ``Gaussian'' are 
 spectral models in the {\it xspec} package; ``wabs'' represents the
 interstellar photoelectric absorption, ``bremss'' represents  bremsstrahlung,
 and ``Gaussian'' represents an iron line. }\\
\multicolumn{8}{p{16cm}}{$\dagger$: The normalization of ``bremss'' is in unit of
 $3.02\times10^{-15}/(4{\pi}D^2)\times\int n_{\rm e} n_{\rm i} dV$ [cm$^{-5}$], where
 $n_{\rm e}$ and $n_{\rm i}$ are electron and ion densities, $V$ is the
 volume, and $D$ is the distance to the object.}
\end{tabular}
\end{center}
\end{table}
\vspace{1cm}

\section{Data Analysis and Results}
\subsection{Spectral Fitting with the Standard Disk Model}
In order to grasp how the X-ray spectra of 4U 1630-47 varied
with time, we fitted all of the spectra with the MCD model plus a powerlaw model, multiplied
by the interstellar photoelectric absorption. We left all of the continuum
parameters free to vary. Since the interstellar absorption toward
4U 1630-47 is not well known, we first fitted all of the spectra with the
column density, $N_{\rm H}$, left free, and then obtained an average value of
$N_{\rm H}$ = $9.5 \times 10^{22}$ cm$^{-2}$. We then repeated the
fitting with $N_{\rm H}$ fixed to this value. 
The $N_{\rm H}$ values taken or obtained in previous studies are in the
range of (5--12)$\times10^{22}$ cm$^{-2}$ 
(Tomsick et al. 1998; \cite{Kuulkers1998}; Cui et al. 2000;
\cite{Dieters}; Trudolyubov et al. 2001), and thus
our value is consistent.
The subsequent analysis employs the same value throughout.
Because a broad iron-K absorption edge is often observed for black-hole
binaries (\cite{Ebisawa1993}), we multiplied the powerlaw
component by a smeared-edge model (hereafter smedge model) 
to improve the fits around the iron-edge region. 
The edge energy, $E_{\rm smedge}$, was typically
found to be around 9 keV, and the optical depth, $\tau_{\rm smedge}$, 
was 0.1--3, with rather large errors.
The edge width was not constrained.
Therefore, hereafter, we do not discuss the edge feature.
Hereafter, we call the above model as an ``MCD plus powerlaw model''.

\begin{figure}[hbt]
\begin{minipage}[tbhn]{8cm}
\begin{center}
\resizebox{8.5cm}{!}{\includegraphics{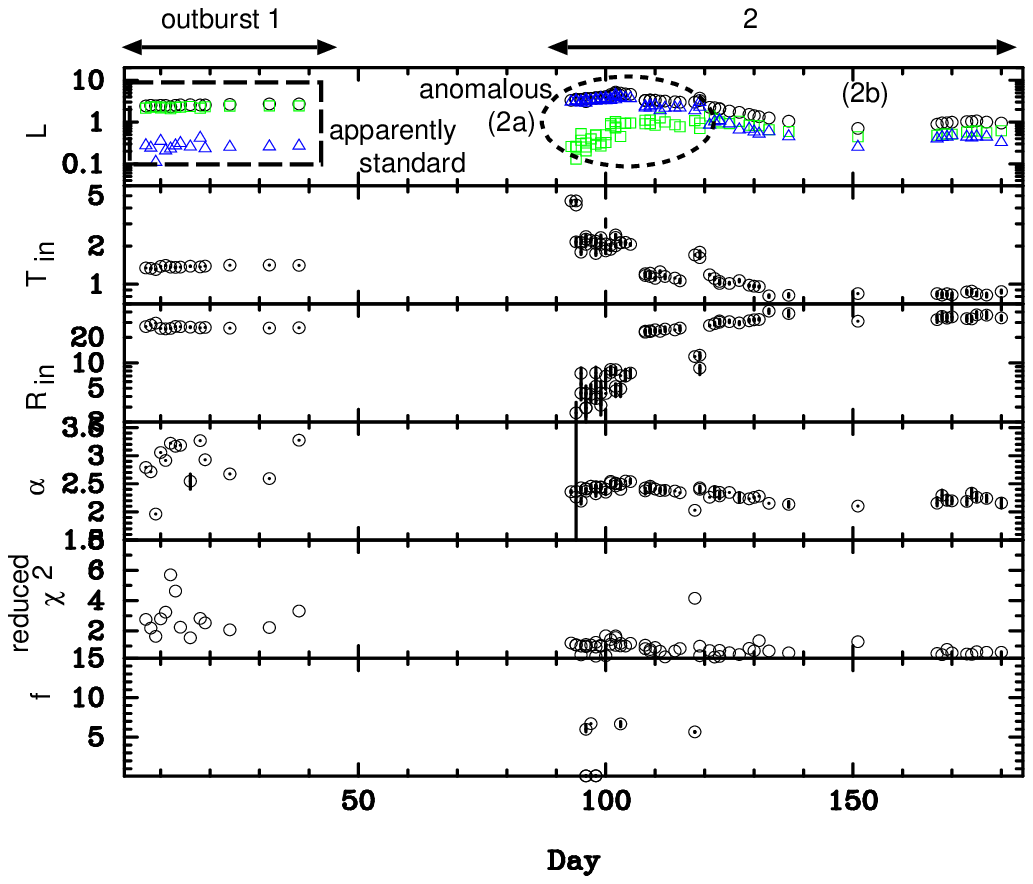}}
\end{center}
\end{minipage}\quad
\begin{minipage}[tbhn]{8cm}
\begin{center}
\resizebox{8.0cm}{!}{\includegraphics{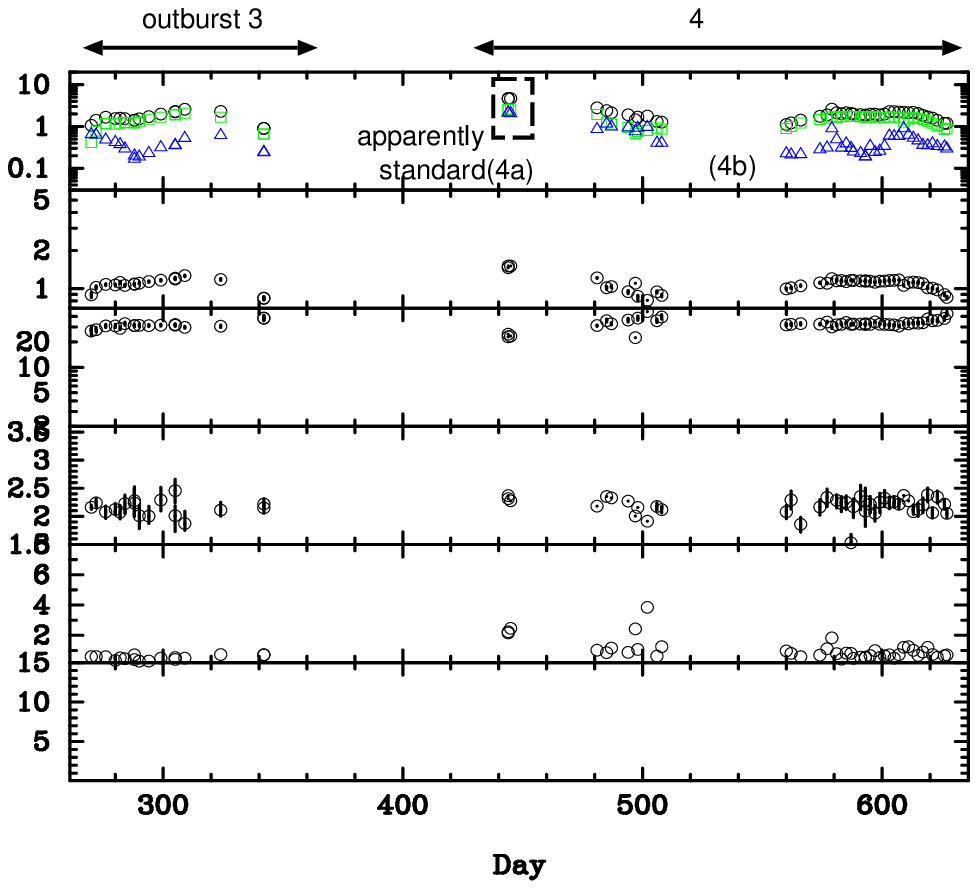}}
\end{center}
\end{minipage}
\resizebox{8.5cm}{!}{\includegraphics{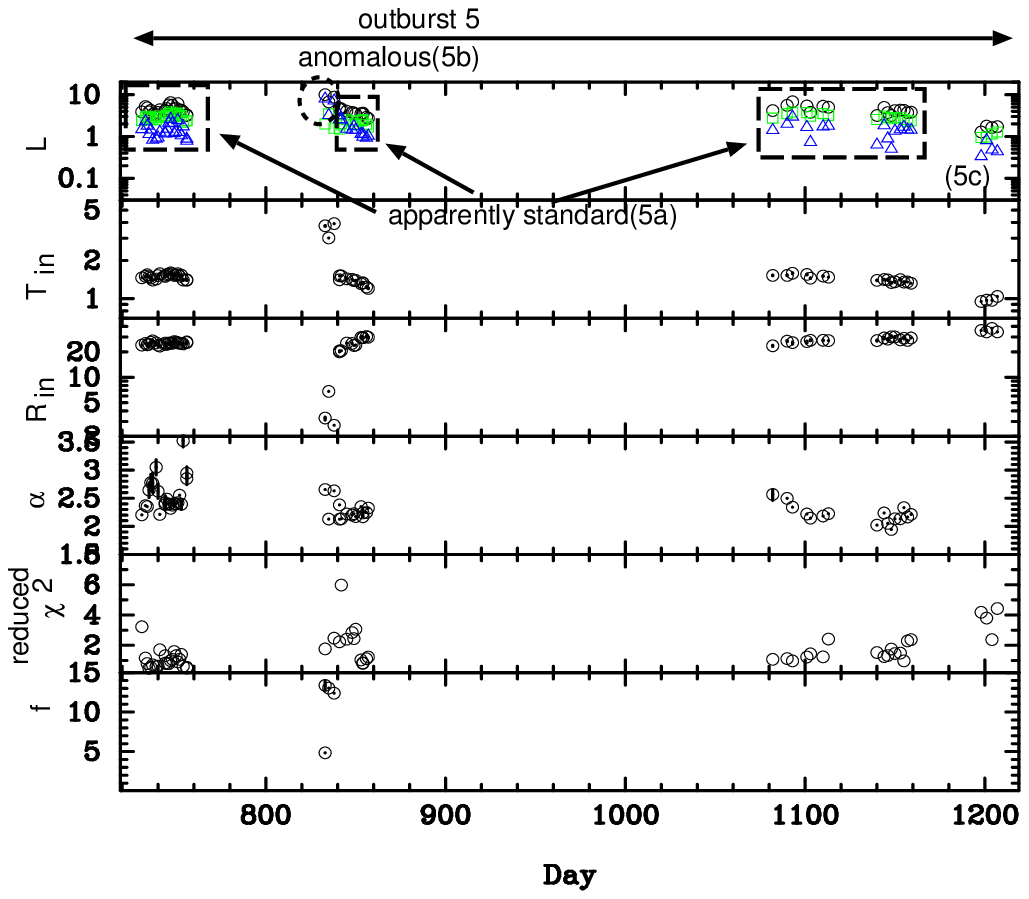}}
\begin{center}
\caption{Time histories of the spectral parameters obtained by fitting
 the PCA spectra
  with an MCD plus powerlaw model (MD model in table 3). 
Only the pointing observations of five
  outbursts are plotted. The light curves of outbursts 1 and 2 are
 shown in the top-left panel, those of outbursts 3 and 4 are in
 the top-right panel, and those of outburst 5 are in the bottom-left panel.
In each panel, time histories of luminosity $L$ [$10^{38}$ erg s$^{-1}$],
 $T_{\rm in}$ [keV], $R_{\rm in}$ [km],
 photon index $\alpha$ of the powerlaw, reduced $\chi^2$, and
 QPO frequency $f$
 [Hz] are plotted from top to bottom.
In the time history of luminosity,
  green squares, blue triangles, and black circles indicate 
  the disk luminosity $L_{\rm disk}$, the powerlaw luminosity 
  $L_{\rm pow}$, and the total luminosity $L_{\rm tot}$, respectively.
The horizontal axis of the plot is defined as an offset day from the
date that is specific for each outburst.
The origin dates are MJD 50200, 50762, 51038, 51420, and 51798 for 
outbursts 1, 2, 3, 4 and 5, respectively.
For reducing the panel number, we plotted the outburst 1 and 2 in a
 single panel, and also did the outburst 3 and 4.
The dashed ellipses and squares indicate the {\it
 anomalous regime} and {\it apparently standard regime}, respectively,
 while others are in the {\it standard regime}.
For the labels ``2a'', ``2b'', ``4a'', ``4b'', ``5a'', ``5b'', and
 ``5c'' , see the text in detail.
} 
\label{lc}
\end{center}
\end{figure}

The best-fit parameters for a typical spectrum observed in each outburst
are given in table 3, where the above model is indicated as ``MD'' model.
In figure \ref{lc}, we show time histories of the X-ray luminosity, 
best-fit
parameters, and reduced $\chi^2$. 
Based on the identification of the soft/hard transition by \citet{Tomsick},
we here do not treat the data with an X-ray luminosity of 
$<6\times10^{37}$ erg s$^{-1}$, since such data are considered to be in the
low/hard state.
In addition, we discarded the data whose inner disk temperature is less than
0.8 keV, because the heavy interstellar absorption does not allow us to 
study such soft emission with confidence.
In previous works, only outburst 2 was analyzed well, and our results
concerning the MCD plus powerlaw model fits in the outburst 2 are almost 
consistent with those in 
\citet{Oosterbroek} and \citet{Tomsick}.

\begin{figure}[hbt]
 \begin{center}
  \resizebox{8cm}{!}{\includegraphics{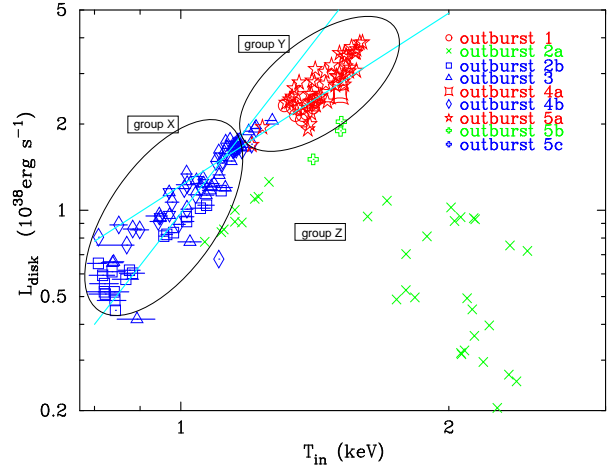}}
\end{center} 
\caption{$L_{\rm disk}$ plotted against $T_{\rm in}$ obtained by the MCD
 plus powerlaw model (MD model). 
The blue, green, and red points represent group X, Y, 
and Z, respectively.
$L_{\rm disk}$ is in units of $10^{38}$ erg s$^{-1}$.
See the text for the definition of the groups. The solid lines
 represent the $L_{\rm disk} \propto T_{\rm in}^4$ and $L_{\rm disk}
 \propto T_{\rm in}^2$ relations, where the disk inner radius 
$R_{\rm in}$ is assumed to be 46.1 km for the former relation.}
\label{lxkt}

\vspace{0.5cm}
\begin{minipage}[tbhn]{8cm}
\begin{center}
(a) Observation A (outburst 1) \\
\qquad {\it apparently standard regime}
  \rotatebox{-90}{\resizebox{5cm}{!}{\includegraphics{psfile/figure4a.ps}}}
 \end{center}
\end{minipage}
\begin{minipage}[tbhn]{8cm}
 \begin{center}
(b) Observation B (outburst 2a) \\
\qquad {\it anomalous regime}
  \rotatebox{-90}{\resizebox{5cm}{!}{\includegraphics{psfile/figure4b.ps}}}
 \end{center}
 \end{minipage}
\begin{minipage}[tbhn]{8cm}
 \begin{center}
\vspace{0.5cm}
(c) Observation C (outburst 2b) \\
\qquad {\it standard regime}
  \rotatebox{-90}{\resizebox{5cm}{!}{\includegraphics{psfile/figure4c.ps}}}
 \end{center}
 \end{minipage}
\begin{minipage}[tbhn]{8cm}
 \begin{center}
\vspace{0.5cm}
\hspace*{1cm}
(d) Observation F (outburst 5a) \\
\qquad {\it apparently standard regime}
\hspace*{1.0cm}
  \rotatebox{-90}{\resizebox{5cm}{!}{\includegraphics{psfile/figure4d.ps}}}
 \end{center}
 \end{minipage}
\caption{Examples of the spectra in different spectral
 states. The data and their 1$\sigma$ errors are indicated by crosses.
The solid line shows the best-fit model (MCD plus powerlaw
 model in the text, or MD model in table 3), which includes an MCD
 (dashed line), a powerlaw (dot-dashed line), and a Galactic
 background (dot line). Bottom graph of each panel shows residuals of the source counts from the best-fit model.}
\label{5}
\end{figure}

The apparent inner disk radius, $r_{\rm in}$, is calculated 
as $K_{\rm MCD}=\left[\left(r_{\rm in}/1\quad{\rm km}\right)\times\left(10\quad{\rm
kpc}/D\right)\right]^2 \cos{\theta}$. 
The value of $r_{\rm in}$ can be related to the true inner radius,
$R_{\rm in}$, as $R_{\rm in}=\kappa^2 \xi r_{\rm in}$ (\cite{Makishima2000}). 
Here, $\kappa$ $\simeq$ 1.7--2.0
(\cite{Shimura}) is a spectral hardening factor, and $\xi$ = 0.41
(\cite{Kubota1998}) is a correction factor for the inner boundary
condition. 
We calculated the disk
luminosity as $L_{\rm disk}=4\pi r_{\rm in}^{2}\times \sigma T_{\rm
in}^4$ from the best-fit apparent inner radius, $r_{\rm in}$, and the inner disk 
temperature, $T_{\rm in}$. 
The powerlaw luminosity, $L_{\rm pow}$, in the 1--100 keV
band is derived from the best-fit model,
assuming an isotropic emission, and then the total luminosity is calculated
as $L_{\rm tot}$ = $L_{\rm disk}$ + $L_{\rm pow}$.

In figure \ref{lxkt}, we show a scatter plot between $L_{\rm disk}$ and $T_{\rm
in}$ in order to see the overall behavior of the X-ray spectra.
The data points, defined as group X in figure \ref{lxkt},
generally follow the relation $L_{\rm disk} \propto T_{\rm in}^4$
(the steeper solid line in the figure) which is expected for a standard
accretion disk with a constant inner radius.
However, two other groups of data deviate from this
relation.
One group (Y) of data shows $L_{\rm disk}$ to be higher than
$1.5\times10^{38}$ erg s$^{-1}$, and
follow a flatter dependence on $T_{\rm in}$ than the standard relation of 
$L_{\rm disk}\propto T_{\rm in}^4$.
The other group (Z) consists of those data points whose $L_{\rm
disk}$ is much lower than those of the
former two groups for the same $T_{\rm in}$, or whose $T_{\rm in}$ is
unusually high for black hole binaries.
Below, we examine the data of individual groups to understand their
spectral states.

Group X includes outburst 3 and
the latter part of outbursts 2, 4, and 5 (hereafter 2b, 4b, and 5c,
respectively).
These data points have a constant $R_{\rm in}$, though $L_{\rm disk}$ changes from $5\times 10^{37}$ to $2 \times10^{38}$ erg s$^{-1}$.
The fits are acceptable, and as shown in figure \ref{5}c and table 3 for
outburst 2b, 
the spectrum is represented by a dominant soft component and a powerlaw 
tail. 
These results indicate that the standard disk picture is realized in these
periods. We hence ascribe the group-X data to the {\it standard regime}, after \citet{Kubota2001} and \citet{Kubota2004}. 
The true inner radius, $R_{\rm in}$, can be 
estimated to be 30--45$\left(D/10\quad{\rm kpc}\right)\left[\cos{\theta}/\left(1/\sqrt{3}\right)\right]^{-\frac{1}{2}}$
km, corresponding to a non-spinning black-hole mass of
3.4--5.1$\left(D/10\quad{\rm kpc}\right)\left[\cos{\theta}/\left(1/\sqrt{3}\right)\right]^{-\frac{1}{2}}\MO$. 
These values are reasonable for a black hole binary.

The former part of outburst 2 and
the middle part of outburst 5 (hereafter, 2a and 5b respectively)
are included in group Z.
Sometimes, the values of $R_{\rm in}$ are less than 10 km 
and $T_{\rm in}$ is also as high as 2 keV or more. 
These values are unusual for black hole binaries.
The typical spectrum in outburst 2a is shown in figure \ref{5}b.
The hard powerlaw component is dominant compared with those
in the $standard$ $regime$ (figure \ref{5}c).
The fits are relatively poor; the reduced $\chi^2$ is relatively large in
outburst 2a and quite large, $>$2, in outburst 5b.
The luminosity of the powerlaw component dominates that of the MCD
component for almost all of the group Z data, as seen in figure \ref{lc}.
Outburst 2a contains the period of the very high state
defined in \citet{Trudolyubov}.
These properties make this outburst reminiscent of the
{\it anomalous regime} of GRO J1655-40 and XTE J1550-564
(Kubota et al. 2001, 2005). 
Hence, these data points are thought to correspond to the {\it anomalous
regime}, which seems to appear at $L_{\rm
tot}\geq2.5\times10^{38}$ erg s$^{-1}$ for 4U 1630-47.

Outburst 1 and the former part of outburst 4 (4a) are 
classified into group Y,
so are two parts (outburst 5a) of outburst 5 before and after
outburst 5b (figure \ref{lc}).
In outburst 1 (figure 4a and table 3), the reduced $\chi^2$ value is
significantly large, even though the spectral shape is represented by the
dominant soft component with a weak powerlaw tail 
as well as the {\it standard regime}. 
Large reduced $\chi^2$ values indicate that the spectra cannot be well
represented by the standard disk model.
As shown in figure \ref{5}a, the residual is significant between the data and
MCD model around 5 keV.
A narrow deep-edge structure is 
required around 7 keV for the powerlaw component in the MCD plus
powerlaw model fitting, 
although this is thought to be artificial
to explain the residual between the data and the MCD model, which is not the
best model.
Throughout this outburst, $L_{\rm disk}$ is very high at
$>1.5 \times 10^{38}$ erg s$^{-1}$, $L_{\rm pow}$ is quite low compared with $L_{\rm tot}$ (figure \ref{lc}), 
and $R_{\rm in}=25$ km is slightly smaller than that in the {\it standard
regime}. 
Furthermore, the photon index of the powerlaw component is larger than
2.5, in contrast to the typical value of 2.0--2.5 
in the {\it standard regime} of this source. 
These features of outburst 1 are the same as found in the {\it apparently standard
regime} of XTE J1550-564. 
Outbursts 4a and 5a show several features that are somewhat 
different from those of outburst 1.
The powerlaw component is not as weak as shown in figure \ref{5}d, and 
the photon index of the powerlaw component, 2.2--3.0, is only slightly
larger than the typical values in the {\it standard regime}.
Nevertheless, the disk luminosity $L_{\rm disk}$ is very high at 
$>1.5 \times 10^{38}$ erg s$^{-1}$, and
follows the relation of $L_{\rm disk}\propto
T_{\rm in}^2$, as well as outburst 1.
Therefore, while the spectral features of the 
powerlaw component are different among outbursts 1, 4a, and 5a, 
the behavior of the MCD component in these periods
is similar to that of the {\it apparently standard regime}.

\begin{table} [hpbt]
\vspace{0.5cm}
\label{para}
\caption{The best-fit parameters. }
\begin{center}    
\hspace*{-1cm}                    
\begin{scriptsize}
\begin{tabular}{ccccccccccc} \hline\hline
ID$^{\ast}$ & \\
model$^{\dagger}$ & $N_{\rm H}$ $^{\ddagger}$ & $T_{\rm in}$ $^{\ddagger}$ & $R_{\rm in}$ $^{\ddagger}$ &
 $\alpha_{\rm ph}$ $^{\ddagger}$ & $N_{\rm thc}$ $^{\ddagger}$ &
 $L_{\rm disk}/L_{\rm thc}/L_{\rm pow}$ $^{\ddagger}$ & $E_{\rm smedge}$
 $^{\ddagger}$ &
 $\tau_{\rm smedge}$ $^{\ddagger}$ & $W_{\rm smedge}$ $^{\ddagger}$ & $\chi^2$/d.o.f. \\ 
& &(keV) & (km) & & $\Gamma_{\rm thc}$ $^{\ddagger}$ &
 (10$^{38}$ erg s$^{-1}$) & (keV) &  & (keV) \\
& & & & & $p$ $^{\ddagger}$ & & &  & \\
 \hline 
\multicolumn{10}{l}{A (outburst 1) : {\it apparently standard regime}} \\
MD & 9.5 & 1.32$^{+0.05}_{-0.01}$ & 28.6$^{+0.7}_{-0.3}$ & 2.71$_{-0.74}^{+0.29}$ & -
 & 2.24/-/0.11 & 7.00$^{+0.52}_{-0.00}$ & $>$1.00& $<$0.01 & 92.7/52 \\ 
PF & 9.5 & 1.36$_{-0.04}^{+0.07}$ & 23.3$_{-4.7}^{+7.9}$ & 2.0(fix) & 0.65$_{-0.08}^{+0.16}$& 
1.74/-/0.34 & 7.00$_{-0.00}^{+0.38}$ & 0.15$_{-0.05}^{+0.12}$ & 0.21$_{-0.21}^{+0.68}$ & 29.7/49 \\ \hline
\multicolumn{10}{l}{B$^{\S}$ (outburst 2a) : {\it anomalous regime}} \\
MD & 9.5 & 1.83$^{+0.11}_{-0.06}$ & 7.15$^{+0.9}_{-1.1}$ & 2.33$_{-0.03}^{+0.03}$ & -
 & 0.52/-/3.21 & 7.76$^{+0.39}_{-0.14}$ & 0.05$^{+0.12}_{-0.03}$ & 10.8$^{+1.7}_{-5.5}$ & 196.6/107 \\
CM & 9.5 & 0.97$^{+0.07}_{-0.03}$ & 41.9$^{+9.9}_{-4.6}$ &
 2.0 (fix) & 1.47$^{+0.21}_{-0.16}$ & 0.57/2.13/0.73 & 9.00$^{+0.0}_{-0.30}$ &
 0.99$_{-0.52}^{+0.37}$ & 28.4$^{+1.6}_{-3.22}$ & 103.2/107 \\
& & & & & 2.40$^{+0.45}_{-0.15}$ & & & & & \\
\hline
\multicolumn{10}{l}{C (outburst 2b) : {\it standard regime}} \\
MD & 9.5 & 0.83$^{+0.02}_{-0.02}$ & 37.3$^{+2.4}_{-2.1}$ & 2.11$_{-0.04}^{+0.04}$ & - & 0.81/-/0.44 & 8.75$^{+0.25}_{-0.22}$ & $>$1.00 & 6.13$^{+0.91}_{-3.47}$ & 49.4/51 \\ \hline
\multicolumn{10}{l}{D (outburst 3) : {\it standard regime}} \\
MD & 9.5 & 1.03$^{+0.02}_{-0.02}$ & 27.6$^{+1.1}_{-1.0}$ & 2.21$_{-0.05}^{+0.06}$ & -
 & 1.18/-/0.48 & 8.92$^{+0.08}_{-0.37}$ & $>$1.00 & 8.92$^{+2.98}_{-4.46}$ &  35.6/46 \\ \hline
\multicolumn{10}{l}{E (outburst 4a) : {\it apparently standard regime}} \\
MD & 9.5 & 1.16$^{+0.01}_{-0.01}$ & 31.6$^{+0.5}_{-0.5}$ & 2.21$_{-0.11}^{+0.10}$ & -
 & 1.66/-/0.44 & 8.84$^{+0.16}_{-0.35}$ & 0.65$^{+0.35}_{-0.29}$ & 2.33$^{+2.16}_{-1.14}$ & 55.2/46 \\
PF & 9.5 & 1.22$_{-0.05}^{+0.05}$ & 26.9$_{-7.4}^{+12.6}$ & 2.0(fix)
 & 0.69$_{-0.14}^{+0.30}$ & 1.44/-/0.34 & 8.99$_{-0.55}^{+0.01}$ &
 0.31$_{-0.13}^{+0.69}$ & 2.27$_{-1.40}^{+6.04}$ & 34.7/50 \\ \hline
\multicolumn{10}{l}{F (outburst 5a) : {\it apparently standard regime}} \\
MD & 9.5 & 1.55$^{+0.01}_{-0.01}$ & 23.6$^{+0.5}_{-0.3}$ & 2.50$_{-0.09}^{+0.09}$ & - & 2.96/-/1.47 & 8.24$^{+0.76}_{-0.53}$ & $>$1.00 & 23.2$^{+6.86}_{-23.2}$ & 64.4/47 \\ 
PF & 9.5 & 1.75$_{-0.05}^{+0.05}$ & 13.2$_{-2.2}^{+1.8}$ &2.0(fix) &0.54$_{-0.03}^{+0.04}$&  1.48/-/1.09 &8.57$_{-0.61}^{+0.43}$ & 0.19$_{-0.09}^{+0.81}$ & 1.69$_{-1.25}^{+10.7}$ &31.6/44 \\ \hline
\multicolumn{10}{l}{G$^{\S}$ (outburst 2a) : {\it anomalous regime}} \\
MD & 9.5 & 1.22$^{+0.02}_{-0.02}$ & 23.5$^{+1.1}_{-0.6}$ & 2.35$_{-0.03}^{+0.04}$ & -
 & 1.04/-/2.02 & 8.98$^{+0.02}_{-0.62}$ & 0.71$^{+0.29}_{-0.32}$ & 29.8$^{+0.16}_{-28.4}$ & 115.5/109 \\
CM & 9.5 & 1.06$^{+0.05}_{-0.03}$ & 40.42$^{+1.8}_{-2.8}$ &
 2.0 (fix) & 1.03$^{+0.18}_{-0.18}$ & 1.14/1.12/0.99 & 9.00$^{+0.0}_{-0.48}$ &
 0.63$_{-0.52}^{+0.37}$ & 5.92$^{+24.33}_{-3.22}$ & 121.3/106 \\
& & & & & 2.73$^{+0.21}_{-0.20}$ & & & & & \\
CM$^{\Vert}$ & 7.64$^{+0.13}_{-0.12}$ & 1.10$^{+0.04}_{-0.05}$ &
 35.1$^{+1.9}_{-1.7}$ & 2.0 (fix) & 0.89$^{+0.17}_{-0.16}$ & 0.96/1.06/1.01 &
 9.00$^{+0.00}_{-0.547}$ & 0.23$^{+0.77}_{-0.13}$ & 5.56$^{+24.44}_{-3.52}$ &
 191.2/210 \\ 
& & & & & 2.70$^{+0.08}_{-0.10}$ & & & & & \\
\hline
\multicolumn{10}{p{16cm}}{$\ast$:'A', 'B', 'C', 'D', 'E', 'F', and 'G' 
corresponds to the observation ID 10411-01-04-00(MJD=50209), 30178-01-10-00(50862), 30172-01-08-00(50899), 40418-01-03-00(51310), 60118-01-07-00(52001), 70417-01-02-01(52532), and 30188-02-20-00(50870).}\\
\multicolumn{10}{p{16cm}}{$\dagger$:``MD'' is a MCD model plus a powerlaw model, multiplied by an interstellar absorption and a 
smeared edge (called as MCD plus powerlaw model in the text). 
``CM'' is a model in which the $thcomp$ model is
 included in addition to the ``MD'' model. ``PF'' is a $p$-free disk
model (see text subsection 4.3) plus a powerlaw model, multiplied by an interstellar absorption and a 
smeared edge. }\\
\multicolumn{10}{p{16cm}}{$\ddagger$:$N_{\rm H}$ is the column density of the
 Galactic interstellar absorption in unit of $10^{22}$ cm$^{-2}$. 
$T_{\rm in}$ and $R_{\rm in}$ are an
 inner disk temperature and radius, respectively. $\alpha_{\rm ph}$
 is a photon index of the powerlaw component. $N_{\rm thc}$ and
 $\Gamma_{\rm thc}$ are a normalization and a spectral index of
 the $thcomp$ model (see text subsection 4.2). $p$ is a parameter of the $p$-free disk model. 
$L_{\rm disk}$, $L_{\rm pow}$, and $L_{\rm thc}$
 are the luminosity of the MCD, powerlaw (1--100 keV), and $thcomp$
 (0.01--100 keV) models, respectively. $E_{\rm smedge}$, $\tau_{\rm
 smedge}$ and $W_{\rm smedge}$ are an edge energy, depth and width,
 respectively, for the smeared edge model.}\\
\multicolumn{10}{p{16cm}}{$\S$:Results by fitting the PCA and HEXTE
 spectra. The data other than 'B' and 'G' are results by using the 
PCA spectrum.}\\
\multicolumn{10}{p{16cm}}{$\Vert$:Results by simultaneously fitting the PCA, HEXTE, and ASCA-GIS spectra.}
\end{tabular}
\end{scriptsize}
\end{center}
\vspace{-0.15cm}
\end{table}

\clearpage

\subsection{Consideration of the Inverse Compton Scattering}
\begin{figure}[hbt]
\begin{minipage}[tbhn]{8cm}
\begin{center}
\resizebox{8.5cm}{!}{\includegraphics{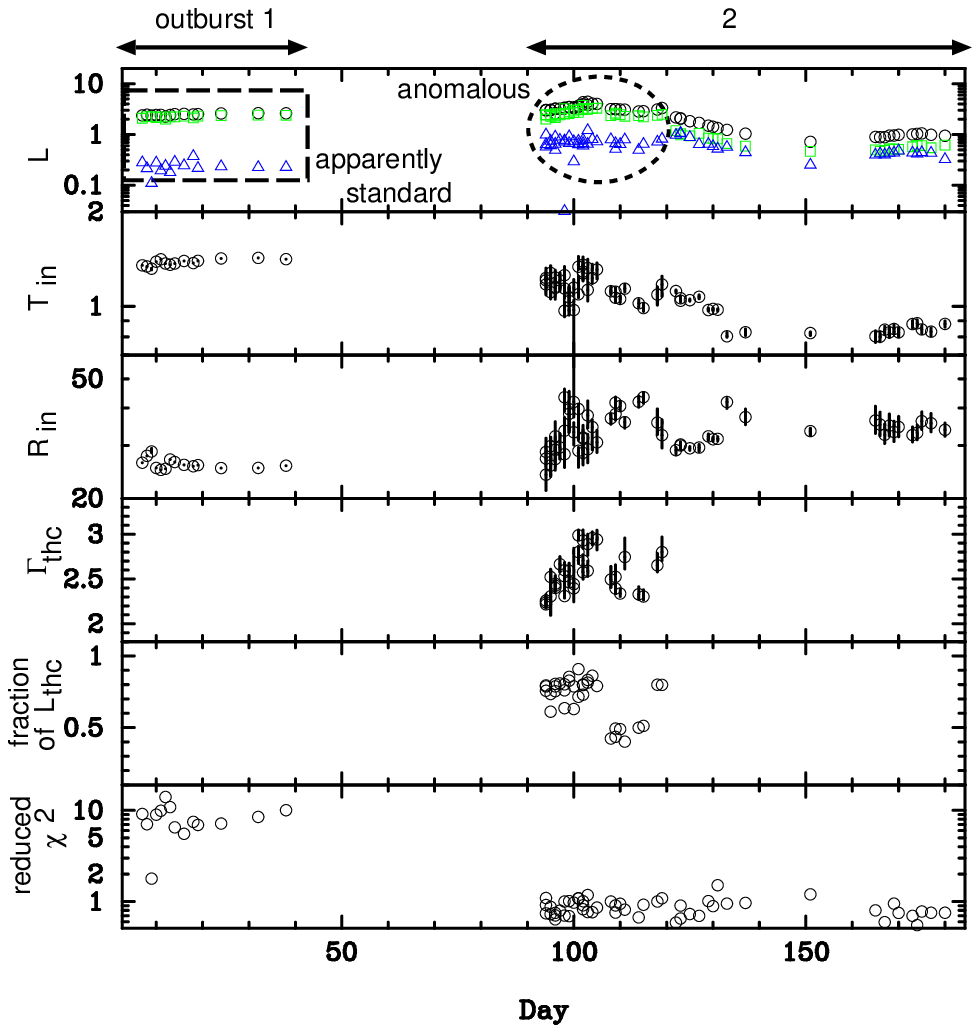}}
\end{center}
\end{minipage}\quad
\begin{minipage}[tbhn]{8cm}
\begin{center}
\resizebox{8cm}{!}{\includegraphics{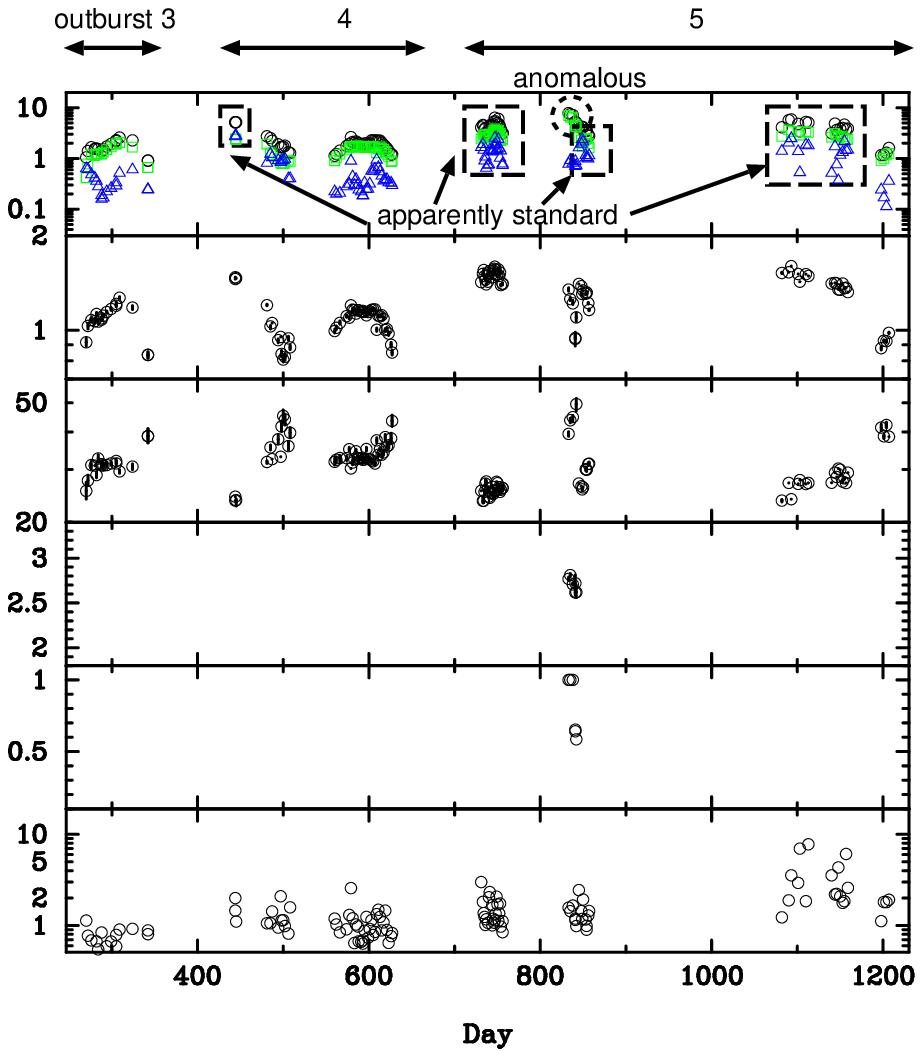}}
\end{center}
\end{minipage}
\caption{Same as figure \ref{lc} concerning the {\it standard regime} and the
 {\it apparently standard regime} data, but results in the {\it
 anomalous regime} are replaced by those obtained with the
 CM model.
In each panel, time histories of luminosity $L$ [$10^{38}$ erg s$^{-1}$],
 $T_{\rm in}$ [keV], $R_{\rm in}$ [km],
 spectral index $\Gamma_{\rm thc}$ of the {\it thcomp}, fraction of
 $L_{\rm thc}$ against $L_{\rm disk}+L_{\rm thc}$, and reduced $\chi^2$ are plotted from top to bottom. The horizontal axis represents the same as that in figure \ref{lc}.}
\label{6}
\end{figure}

In the group-Z data, 
the spectral features are in good agreement with
those in the $anomalous$ $regime$ of GRO J1655-40 and XTE J1550-564
(Kubota et al. 2001, 2005; \cite{Kobayashi}). We therefore
consider, after these previous studies, that the enhanced powerlaw component
with a steep slope is produced when a certain fraction of photons from
the optically-thick disk are Compton-up-scattered by hot
electrons around the disk. 
Following \citet{Kubota2004}, we
refitted these spectra with a three-component model,
in which a thermal Comptonization component ({\it thcomp};
\cite{Zycki}) is added to the original MCD plus powerlaw model. 
We hereafter refer this model to the ``CM'' model.
We assumed that the seed photons are provided by the MCD emission, so the maximum color temperature of the seed photons was
tied to $T_{\rm in}$. 
Furthermore, the electron temperature, $T_{\rm e}$, was fixed to 20 keV.
Hence, the $thcomp$ model has two free parameters:
a $thcomp$ photon index, $\Gamma_{\rm thc}$, which expresses the spectral
shape below $kT_{\rm e}$, and a normalization, $N_{\rm thc}$. 
Since the parameters of the $thcomp$ and the powerlaw models can
not be constrained independently, we fixed the photon index of the 
powerlaw component to 2.0, which is the average value in the {\it
standard regime}.
In figure \ref{6}, we show the time histories of the spectral parameters
in the same manner as figure \ref{lc}. 
Thus, the fit has been significantly improved, as shown in table 3
(see the line of ``CM'' model in table 3). 

\begin{figure}[hbpt]
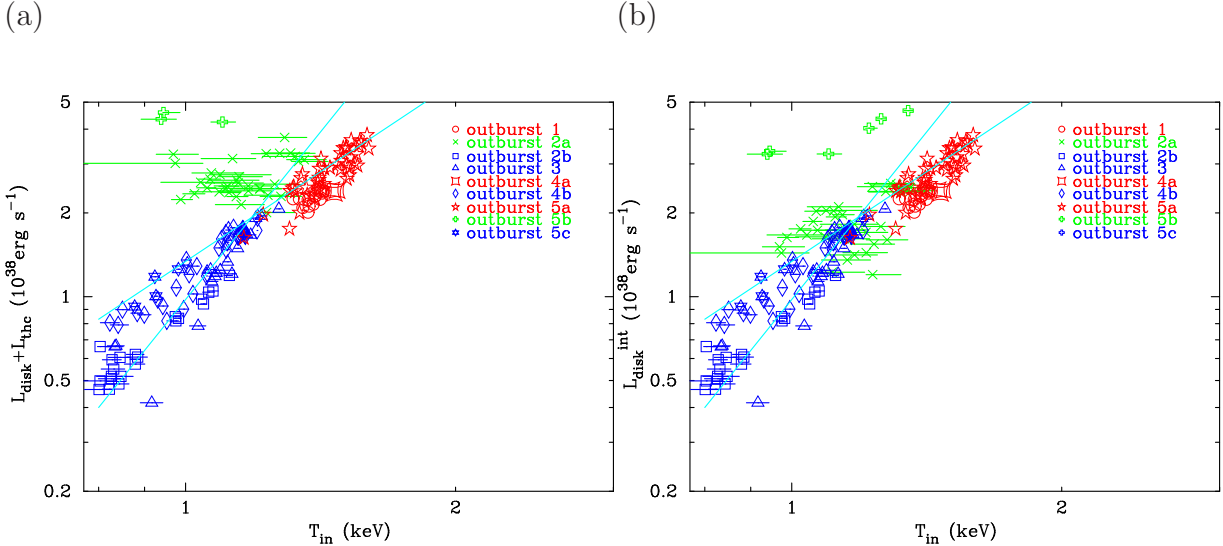

\begin{minipage}[tbhn]{8cm}
(a)
\vspace{-0.5cm}
\begin{center}
  \rotatebox{-90}{\resizebox{6cm}{!}{\includegraphics{psfile/figure6a.ps}}}
\end{center}
\end{minipage}
\begin{minipage}[tbhn]{8cm}
(b)
\vspace{-0.5cm}
\begin{center}
  \rotatebox{-90}{\resizebox{6cm}{!}{\includegraphics{psfile/figure6b.ps}}}
\end{center}
\end{minipage}
\caption{Same as figure \ref{lxkt}, but the data points in the
 {\it anomalous regime} are re-calculated as (a) $L_{\rm disk}+L_{\rm thc}$
 or (b) $L_{\rm disk}^{\rm int}$, where $L_{\rm disk}^{\rm int}$ is
 estimated by considering the conservation of the photon number.
The longitudinal axis is in unit of $10^{38}$ erg s$^{-1}$.
}
\label{t-ldisk}
\end{figure}

Following Kubota, Makishima (2004), we plot $L_{\rm disk}$ + $L_{\rm thc}$
against $T_{\rm in}$ in figure \ref{t-ldisk}a, where $L_{\rm thc}$ is a
luminosity of the {\it thcomp} model in the 0.01--100 keV.
The values of T$_{\rm
in}$ become 0.9--1.4 keV, which are almost intermediate between the
highest end of the {\it standard regime} and the lowest end of the
{\it apparently standard regime}. 
However, the obtained values of
$L_{\rm disk}$ + $L_{\rm thc}$ must be higher than the intrinsic luminosity
of the underlying optically thick disk, $L_{\rm disk}^{\rm int}$, due to the
inverse Compton scattering. 
To correct this effect, following the equation by Kubota, Makishima
(2004), we calculate the intrinsic disk flux as
\begin{eqnarray}
F_{\rm disk}^{\rm p}+F_{\rm thc}^{\rm p}\times 2\cos{\theta}=0.0165\times \left[\frac{r_{\rm in}^2\cdot \cos{\theta}}{\left(D/10\quad{\rm kpc}\right)^2}\right]\times \left(\frac{T_{\rm in}}{1\quad{\rm keV}}\right)^3 \hspace{0.3cm} {\rm photons}\ {\rm s^{-1}}\ {\rm cm^{-2}},
\end{eqnarray}
where $F_{\rm disk}^{\rm p}$ and $F_{\rm thc}^{\rm p}$ are 0.01--100 keV
photon flux of the direct disk component and the Comptonized
component, respectively. 
$T_{\rm in}$ refers to the disk temperature
obtained by considering the inverse Compton scattering. 
The intrinsic disk luminosity, $L_{\rm disk}^{\rm int}$, can be estimated 
as $L_{\rm disk}^{\rm int}=4\pi r_{\rm in}^2\times\sigma T_{\rm in}^4$,
where $r_{\rm in}$ is given by equation (1).
In figure \ref{6}, we plot a time history of $R_{\rm in}$ estimated 
from $r_{\rm in }$ obtained by this relation. 
Almost values of $R_{\rm in}$ are obtained as 30--50 km, which are
in good agreement with those in the {\it standard regime}. 
When the powerlaw photon index is fixed to a larger value of 2.3, 
the values of $R_{\rm in}$ become smaller in such a way from $\sim$40 km
to $\sim$32 km; the difference is not significant, and values
obtained here are both close to that of the standard regime 
compared with the value in the MCD plus powerlaw model fitting.
Figure \ref{t-ldisk}b shows the
estimated $L_{\rm disk}^{\rm int}$ against $T_{\rm in}$; the {\it
anomalous regime} data points now align better along the standard disk
locus, implying that the $R_{\rm in}$ returns to its proper value.
Thus, in the {\it anomalous regime}, the disk itself stays still in the
standard state, whereas the
inverse Compton-scattering process converts a significant fraction of
its emission into the hard component, as suggested in GRO J1655-40 and 
XTE J1550-564.

Even after these corrections, some data points in figure \ref{t-ldisk}b
deviate from
the standard relation. They are from outburst 5b, where the total luminosity becomes
the highest among the {\it RXTE} observations. We speculate that the
underlying disk, itself, becomes deviated in these period from the
standard picture.

\begin{figure}[hbt]
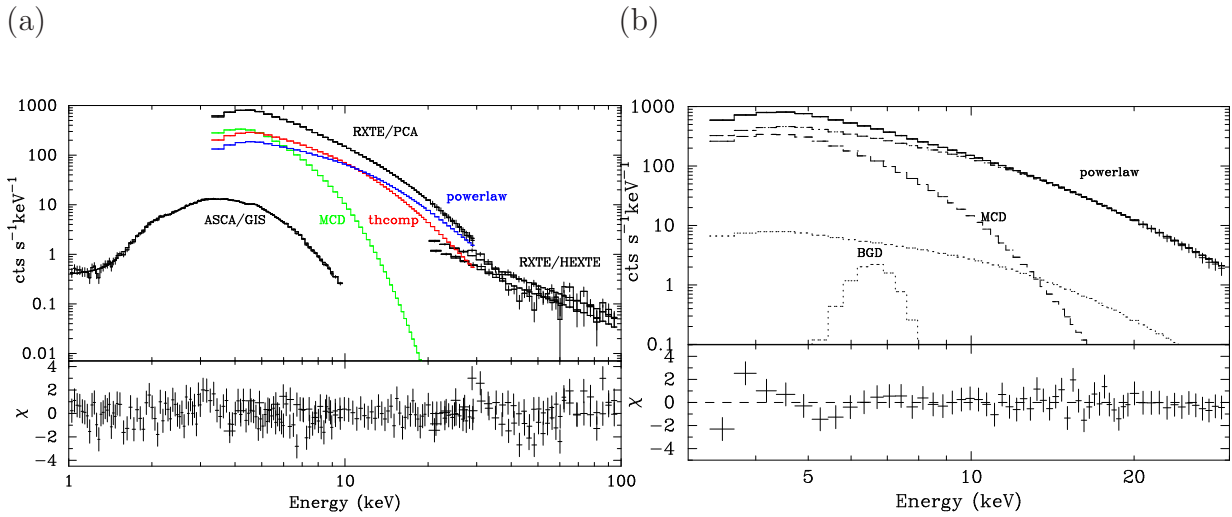

\begin{minipage}[tbhn]{8cm}
(a)
\vspace{-0.5cm}
\begin{center}
  \rotatebox{-90}{\resizebox{5.5cm}{!}{\includegraphics{psfile/figure7a.ps}}}
\end{center}
\end{minipage}
\begin{minipage}[tbhn]{8cm}
(b)
\vspace{-0.5cm}
\begin{center}
  \rotatebox{-90}{\resizebox{5.5cm}{!}{\includegraphics{psfile/figure7b.ps}}} 
\end{center}
\end{minipage}
\caption{(a) Joint fit to the GIS, PCA, and HEXTE 
spectra obtained simultaneously on MJD=50870, with the ``CM'' model. 
For the PCA spectra, the best-fit model and individual three additive 
components (MCD, {\it thcomp}, and powerlaw) are plotted.
Only the best-fit model is shown for the GIS and HEXTE spectra. 
(b) Joint fit to the PCA and HEXTE spectrum on MJD=50870, with the MCD
 plus powerlaw model. Here, for simplicity, only the PCA spectrum is shown.
 The solid, dashed, dot-dashed, and dot lines, together with
 cross points, are the same as figure 4. 
}
\label{77}
\vspace{0.5cm}
\end{figure}

In outburst 2a, we have a simultaneous ASCA GIS
coverage at MJD = 50870. 
The GIS covers the lower energy range of 0.7--10 keV than the
PCA, which covers the energy range of 2--60 keV. 
Therefore, we can obtain
more accurate values of $T_{\rm in}$ and $N_{\rm H}$ by using the GIS
data. 
We included the 1--10 keV GIS spectrum in addition to the PCA and HEXTE spectra,
and fit them simultaneously with the above ``CM'' model, 
without fixing the value of $N_{\rm H}$. 
The results are given in table 3 (label ``G''), and the achieved fit is shown in figure \ref{77}. 
In spite of the wide energy range (1--100 keV) and the extremely high
data statistics (of the GIS and the PCA/HEXTE), the ``CM'' model has
provided an amazingly successful ($\chi^2/$d.o.f = 191.2/210) joint fit to
the spectra. The best-fit parameters have remained close to those
obtained with the {\it RXTE} data alone, without any knowledge for energies below
$\sim$ 3 keV. These results give a strong justification to the
``CM'' modeling. As can be seen from figure \ref{77}a, 
the flux in an
intermediate energy range (5--10 keV) is now carried by the $thcomp$
component, with the MCD and powerlaw components dominating in the lower and
higher energies, respectively. This makes a contrast to the conventional
MCD plus powerlaw fit to the same PCA spectrum (figure \ref{77}b), 
in which
the model tried to reproduce the intermediate-range flux by increasing
$T_{\rm in}$ (which in turn reduced $R_{\rm in}$) and steepening the
powerlaw.

\clearpage
Here, for simplicity, only the PCA spectrum is shown.

\subsection{Evidence of Slim Disk}
The spectral features in outbursts 1, 4a, and 5a (group Y) 
apparently resemble those in the {\it standard regime}.
However, these data points follow the relation $L_{\rm
disk}\propto T_{\rm in}^2$, as noticed in subsection 3.1, 
and the spectral shape
is not necessarily represented by the standard MCD plus powerlaw model.
This suggests that the spectral state in group Y 
corresponds to the {\it apparently standard 
regime} (\cite{Kubota2004};\cite{Kubota2005}), in which the inner
disk may be described by the slim-disk solution.
In order to examine whether outburst 1 data can be described by the 
slim-disk picture, 
we here employ the $p$-free disk model, which mathematically extends
the MCD model (\cite{Mineshige}; \cite{Hirano}; \cite{Kubota2005}). 
This model assumes the radial temperature profile to be as 
$T(r)=T_{\rm in}\left(r/r_{\rm in}\right)^{-p}$, 
where $p$ is a free parameter. 
The profile with $p$ = 0.75 corresponds to the MCD model, while
$p<$0.75 yields a spectrum
softer than the MCD model for the same $T_{\rm in}$. 
That is, a small $p$ suggests a low radiative efficiency at the
inner portion of the accretion disk.

\begin{figure}[hpbt]
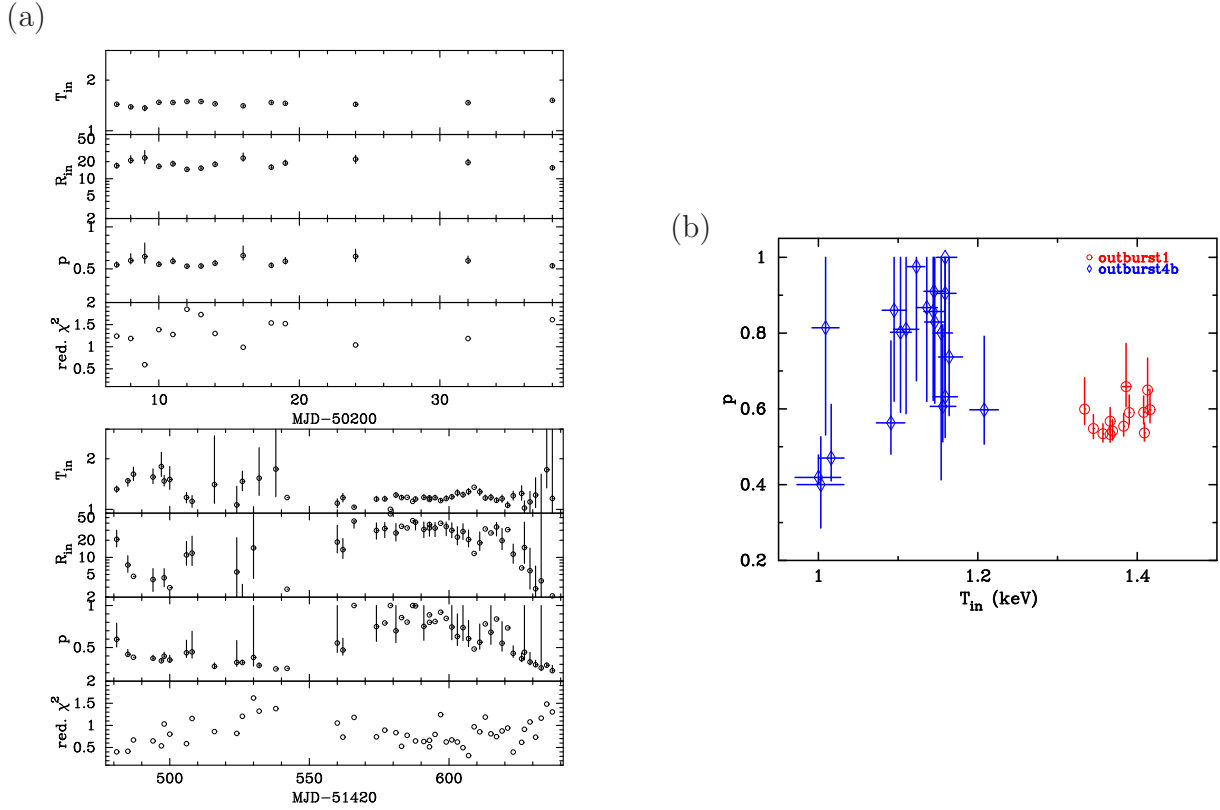

\begin{minipage}[tbhn]{8cm}
(a)
\vspace{-0.5cm}
\begin{center}
 \rotatebox{-90}{\resizebox{5cm}{!}{\includegraphics{psfile/figure8a1.ps}}}
 \rotatebox{-90}{\resizebox{5cm}{!}{\includegraphics{psfile/figure8a2.ps}}}
 \end{center}
\end{minipage}\qquad
\begin{minipage}[tbhn]{8cm}
(b)
\vspace{-0.5cm}
  \rotatebox{-90}{\resizebox{5cm}{!}{\includegraphics{psfile/figure8b.ps}}}
\begin{center}
 \end{center}
\end{minipage}
\caption{(a)Time histories of the best-fit parameters and reduced $\chi^2$
 for the ``PF'' model fits. The top and bottom panels are for outbursts
 1 and 4b, respectively.
(b)Best-fit values of $p$ in the $p$-free disk model, 
 plotted against $T_{\rm in}$ obtained through a spectral fitting with the
 MCD plus powerlaw model. The circles and diamonds correspond to the {\it
 apparently standard regime} and {\it standard regime}, respectively.}
\label{9}
\end{figure}

Following \citet{Kubota2004}, we fitted the spectra of the {\it
apparently standard regime},
by replacing the MCD component with the $p$-free disk model, by fixing 
the powerlaw photon index to 2. 
We hereafter refer this model to the ``PF'' model.
The free parameters are $T_{\rm in}$, $p$, normalization of the $p$-free disk
and the powerlaw model, and the smeared-edge model parameters.
Here, we analyze the data of outbursts 1 and 4b; the former is in the
{\it apparently standard regime}, while the latter is in the {\it
standard regime} for comparison.
The best-fit values and the reduced $\chi^2$ are plotted in figure \ref{9}a.
It can be seen that the reduced $\chi^2$ for outburst 1 
({\it apparently standard regime})
dramatically becomes much smaller from $\sim4$ to $\sim1.5$
(see also figure \ref{lc} and table 3).
The reduced $\chi^2$ values are still somewhat large, 
probably due to the incompleteness of the $p$-free model, which
does not take into account the inner boundary condition and relativistic
effects.
For outburst 4b ({\it standard regime}), the $R_{\rm in}$ values
widely scatter with large errors, indicating that the $p$-free disk model
introduces an extra parameter.
Figure \ref{9}b shows the best-fit values of $p$, plotted against 
$T_{\rm in}$ obtained with the MCD plus powerlaw model fit. 
In this diagram, the values of $p$ in the
{\it apparently standard regime} are 0.5--0.7, which is 
smaller than $\sim$0.8 in the {\it standard regime} at 
$T_{\rm in}\sim1.1$ keV.
Therefore, we suggest that the radiative efficiency at the inner
accretion disk is lower in the {\it apparently standard regime}
than in the {\it standard regime}; another cooling process, like 
advective cooling, is required in addition to radiative cooling. 
The theoretical solution of the slim disk takes into account the effect
of advective cooling, and predicts smaller values of $p$
than the standard disk (e.g., \cite{Watarai}). 
In addition, a relation of $r_{\rm in} \propto T_{\rm in}^{-1}$ is
found, when the spectra predicted by the slim
disk solution are fitted with the MCD plus powerlaw model (\cite{Watarai}).
This explains the relation of $L_{\rm disk}\propto T_{\rm in}^2$ of group Y.
Thus, the X-ray spectra in group Y show evidence of a slim disk. 
This is the second example of such a state after XTE J1550-564.
We note that the value of $p$ at a lower $T_{\rm in}$ in outburst 4b 
({\it standard regime}) is somewhat smaller, 
and such a trend is reported to be due to because
the MCD model is a mere approximation of the exact standard-disk
solution (\cite{Kubota2004, Kubota2005}).

\clearpage

\section{Discussion}

\subsection{Three Spectral States Concerned with the Disk Structure}

Through an extensive spectral analysis of the {\it RXTE} data covering the five
outbursts in 1996--2004, we confirmed that most of the X-ray spectra in 
the high state of 4U 1630-47 can be classified into three 
distinct spectral states.
One state is well explained by the standard accretion disk, 
which appears when $T_{\rm in}$ is less than 1.2
keV, or the X-ray luminosity is below $2.5\times10^{38}$ erg s$^{-1}$. 
In this {\it standard regime}, the data
satisfies the $L_{\rm disk} \propto T_{\rm in}^4$ relation, and thus 
this regime corresponds to the standard high/soft state.
In this section, we give answers to aims 1, 3, and 4 of section 1.

When the X-ray luminosity exceeds $2.5\times10^{38}$ erg s$^{-1}$, the
other two states appear.
One of them, the {\it anomalous regime}, is characterized by the 
dominant steep ($\Gamma \sim$
2.5) powerlaw, as well as the unusual values of $R_{\rm in}$ and $T_{\rm
in}$, which are obtained when fitting the spectra with the MCD plus 
powerlaw model.
The dominant powerlaw component can be understood in such a way
that a significant fraction of the disk emission (soft component) is
converted into the hard component through inverse Compton scattering. 
The apparent variability of $R_{\rm in}$ can be explained by this
effect, and $R_{\rm in}$ remains constant when we consider the
Comptonized photons for the disk flux.
Thus, after GRO J1655-40 and XTE J1550-564, 4U 1630-47 is the third
source to exhibit the {\it anomalous regime}.
The data of this regime is mostly in the very high state or
intermediate state.
Therefore, our interpretation by the Comptonization picture indicates
that, in some parts of the very high state or intermediate state,
the disk structure is still standard, but the ambient hot
electron corona that Comptonizes the disk emission is formed and evolved.
The quantification of spectral structures in the framework of
\citet{Kubota2001} is important, since the black hole mass measurement
is still available, even in the very high state, and 
we can constrain the solid angle of the hot electron corona around the disk
from the fraction of Comptonized photons in the disk emission.

The other state appears in outbursts 1, 4a, and 5a, where the spectral
features are apparently similar to those in the {\it standard regime}.
However, the disk luminosity is found to follow
the relation $L_{\rm disk} \propto T_{\rm in}^2$, indicating
a low radiation efficiency of the accretion disk.
This feature is similar to those of the {\it apparently standard regime} 
observed in XTE J1550-564 (\cite{Kubota2004}), and maybe interpreted
as the formation of an optically thick advective disk, namely, a slim-disk.
The {\it apparently standard regime} also corresponds to the standard
high/soft state with very high X-ray luminosity.
However, X-ray spectra are not exactly represented by the standard MCD
plus powerlaw model, and prefer the $p$-free disk model;
and thus, we here stress that the disk structure 
somewhat deviates from the standard disk.

As above, we can mostly understand the disk structure in the high state of 4U
1630-47 within the framework of \citet{Kubota2001} and \citet{Kubota2004}, after 
GRO J1655-40 and XTE J1550-564.
However, some parts of the spectra cannot be well explained by the above
three spectral states, and thus further studies are needed to understand
the disk structure of a very high state or an intermediate state completely.

\subsection{Flux Fraction of the Powerlaw Component and Critical Luminosity}

\begin{figure}[hbt]
\vspace{0.5cm}
 \begin{center}
  \rotatebox{-90}{\resizebox{7cm}{!}{\includegraphics{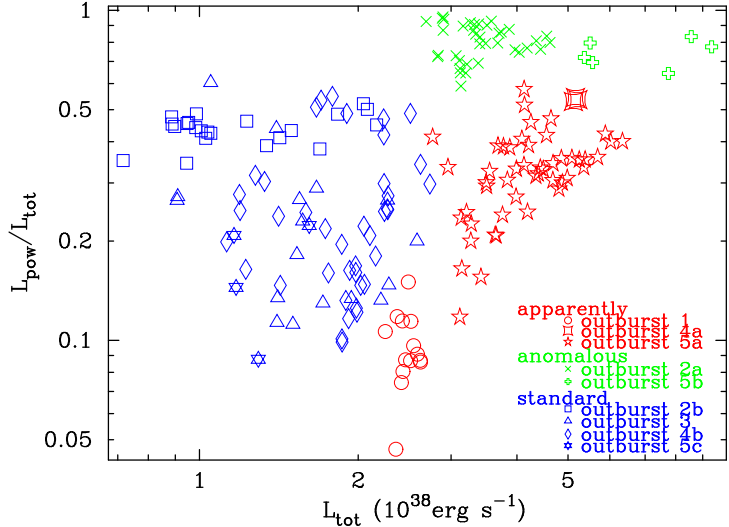}}} 
 \end{center}
\caption{Luminosity ratio of the powerlaw to the total component
  obtained with the MCD plus powerlaw fit, plotted
  against the total luminosity. The luminosities are in units of 
$10^{38}$ erg s$^{-1}$, and refer to values in figure \ref{lc}. 
}
\label{ralx}
\vspace{0.5cm}
\end{figure}

Although we have utilized luminosity-dependent spectral changes to
distinguish the three states, there can be a simpler way of
classification, which is usable even in limited snap-shot observations. 
In figure \ref{ralx}, 
we plot the luminosity ratio of the powerlaw
to the total component, $L_{\rm pow}/L_{\rm tot}$, 
against the total luminosity, $L_{\rm tot}$, where $L_{\rm pow}$ is that
obtained in the MCD plus powerlaw model fitting.
It can be seen that the three states are clearly separated in this
diagram.
The {\it standard regime} (blue points) locates at $L_{\rm
tot}<2.5\times10^{38}$ erg s$^{-1}$.
The {\it anomalous regime} (green points) 
is characterized by $L_{\rm tot}>2.5\times10^{38}$ erg
s$^{-1}$ and $L_{\rm pow}/L_{\rm tot}>0.5$.
The region of $L_{\rm tot}>2.5\times10^{38}$ erg s$^{-1}$ and 
$L_{\rm pow}/L_{\rm tot}<0.5$ corresponds to the {\it apparently 
standard regime} (red points).
The threshold luminosity of $2.5\times10^{38}$ erg s$^{-1}$ should be
different among black hole binaries, and hence cannot be constrained from
the limited data.
However,
the {\it anomalous regime} can be identified only by $L_{\rm pow}/L_{\rm
tot}$.
Therefore, if the data include the {\it anomalous regime},
we can constrain the threshold luminosity, and 
thus identify the {\it apparently standard regime}.
This suggestion can be tested by using the {\it RXTE} data 
of a number of black hole binaries.

Throughout the five outbursts of 4U 1630-47, we have found that the
well-defined critical luminosity of $L_{\rm crit}$ $\sim$ 2.5 $\times$
10$^{38}$ erg s$^{-1}$ divides the {\it standard regime} from the other
two, with a good reproducibility.
This is the first case where the critical luminosity $L_{\rm crit}$ is
almost the same among outbursts of one object, 
indicating that $L_{\rm crit}$ is proper to the object.
Therefore, objective 2 in section 1 is achieved as above.
We may also estimate the Eddington luminosity as $L_{\rm E}$ $\sim$
(6.8--10.2) $\times$ 10$^{38}$ erg s$^{-1}$, using the black hole mass 
estimate of
3.4--5.1 $\MO$ (subsection 4.1). Then, we find $L_{\rm crit}$/$L_{\rm E}$ =
0.25--0.35 for 4U 1630-47. 
This value is somewhat higher than the
values of $L_{\rm crit}$/$L_{\rm E}$ $\sim$ 0.15--0.20 observed from GRO
J1655-40 (Kubota et al. 2001) and XTE J1550-564 (\cite{Kubota2004}),
while lower than that for LMC X--3, $L_{\rm crit}$/$L_{\rm E}$ $>$ 0.9
(Kubota et al. 2001). 
The critical temperature corresponding to $L_{\rm crit}$ is $\sim$1.2 keV for
4U 1630-47 and GRO J1655-40, and 1.0 keV for XTE J1550-564.
Therefore, the transition is not determined only by the temperature.
We speculate that $L_{\rm crit}$ and the transition temperature may be determined by the combination of the black hole mass,
spin, and the accretion rate.

\subsection{Issues on the {\it anomalous regime}}

Since the {\it anomalous regime} is important to understand the unified
view of the very high state or intermediate state, we here briefly
discuss two unresolved issues concerning the {\it anomalous regime}.

Outburst 2 was accompanied by an optically thin 
radio flare (\cite{Hjellming}).
The radio flare lasted from MJD 50840 to MJD 50880, in good agreement
with the {\it anomalous regime} period (figure \ref{lc}).
XTE J1550-564 also exhibits an optically thin radio
flare (\cite{Wu}) in the {\it anomalous regime} (\cite{Kubota2004d}).
In addition, 
for GRO J1655-40, XTE J1550-564, and 4U 1630-47, the radio flare and QPOs
are associated with only the {\it anomalous regime}.
These results 
suggest that the optically thin radio flare and QPOs have relations with 
the Comptonizing plasma, rather than the accretion disk, and 
these two phenomena are thus good indicators of the 
{\it anomalous regime}.
Further studies of the correlation between the {\it anomalous regime} and
a radio flare are interesting from the view point of the disk/jet connection.

In figure \ref{t-ldisk}b, most data points of outburst 5b, which we
identified as the {\it anomalous regime}, deviate from other data points toward
higher luminosity.
Such a deviation is similar to the case of the strong very high state in
XTE J1550-564 \citep{Kubota2004}.
They indicated that the accretion disk is not in the standard state 
and/or does not reach the last-stable orbit.
In the case of 4U 1630-47, the high state continued just before outburst
5b, and thus the situation is different; the truncation of the disk is
possibly caused by extremely high radiation pressure.
Since the total luminosity was the highest in this outburst, the
radiative pressure became highest. 
Then, the inner disk material is thought to produce 
a hot and optically thin corona.
As a result, the disk is truncated and the inner disk radius, $R_{\rm
in}$, becomes large, leading to the higher luminosity.

\bigskip

The authors give thanks to an anonymous referee for very useful comments and
careful reading, and to the {\it RXTE} team for the satellite calibration and 
operations.
This work has been supported by the
Grant-in-Aid from the Japanese Ministry of Education, Culture, Sports, Science and
Technology (14079206).

\vspace*{2cm}

\end{document}